# Magnetoelectric interactions in polycrystalline multiferroic antiferromagnets $CuFe_{1-x}Rh_xO_2$ (x=0.00 and x=0.05)

B. Kundys,[ǂ] A. Maignan, D. Pelloquin, Ch. Simon

*Laboratoire CRISMAT, UMR 6508 CNRS/ENSICAEN, 6 bd du Maréchal Juin, F-14050 CAEN Cedex 4, Normandy 14050, France*

Magnetoelectric coupling in the polycrystalline antiferromagnets $CuFe_{0.95}Rh_{0.05}O_2$ and $CuFeO_2$ has been investigated. For both samples, electric polarization was observed in the absence of an applied external magnetic field demonstrating that for multiferroic research ceramics are worth to be studied. The observed magnetodielectric effect for $CuFe_{0.95}Rh_{0.05}O_2$ in the electrically polar phase supports the existence of a noncollinear antiferromagnetic state. Interestingly, the electric polarization of this sample can be suppressed by a magnetic field. The temperature dependence of the relative magnitude of the magnetodielectric effect shows a discontinuity, clearly indicating different mechanisms of the magnetodielectric couplings in polar and paraelectric antiferromagnetic states.

## 1. Introduction

The realization of bulk magnetic multiferroic materials is a challenging problem for chemists, because ferroelectricity and (anti)ferromagnetism are exclusive properties in most solids [1,2]. Magnetoelectric materials, however, could have practical applications, considering the possible magnetic control of electric polarization (or vice versa) for memory or magnetic sensors. The discovery of magnetically induced polarization in noncollinear (spiral) antiferromagnets [3,4] has spurred renewed interest in frustrated antiferromagnetic compounds. Magnetic frustration is responsible for the lack of stable collinear antiferromagnetic structures that favor the spiral configuration. Such magnetic structures have been reported to be able to create electric polarization [5,6]. Among the recently discovered magnetoelectric materials, transition metal oxides provide many of the compounds as for instance $CoCr_2O_4$[7], $MnWO_4$[8,9], CuO[10] and $Ba_2Mg_2Fe_{12}O_{22}$[11]. Of particularly interest are the results on the magnetic field-induced 180° degrees polarization rotation in $CoCr_2O_4$ and $Ba_2Mg_2Fe_{12}O_{22}$[7,11] and the electric polarization flop from one crystallographic direction to another that has been found to occur in rare earth manganites $TbMnO_3$, $DyMnO_3$, $GdMnO_3$ [3,12] and in $MnWO_4$ [8,13]. Because the electrical polarization in these materials lies in the plane of the spin spiral but in a direction that is perpendicular to its propagation vector, most studies have been devoted to crystals to demonstrate magnetic field-induced switching of electric polarization. Taking into account the complexity that is linked to crystal growth and cutting along the desired crystallographic directions, it is very interesting to study polycrystalline materials. With this aim in view, the recent reports on the $CuFeO_2$ delafossite have drawn our attention to this compound [14–22]. This compound crystallizes in a very anisotropic structure that is composed of $FeO_2$ planes in which the high spin $Fe^{3+}$ cations (S=5/2) lie at the center of edge-shared octaheda where other cation can also be substituted (Fig. 1a). The magnetic $Fe^{3+}$ species form a planar triangular network in which the successive FeO2 planes are separated by $Cu^+$ (S=0) cations in a dumbbell-shaped coordination along the stacking direction [23,24]. For $CuFeO_2$ crystals, a complex magnetic diagram was reported evidencing several spin-flop transitions characterized by a region of (H,T) plane with electric polarization linked to a noncollinear incommensurate antiferromagnetic structure [16,17]. Furthermore, this magnetic and electric phase diagram was found to be very sensitive to small doping at the Fe site [18,20] so that for crystals $CuFe_{0.98}Al_{0.02}O_2$, an electric polarization is induced even in the absence of applied external magnetic field [18,21]. For instance, at 2 K, a magnetic field in the range of 7T< H< 13T is needed to induce an electric polarization in pure $CuFeO_2$ [17], whereas the polar state already exists in 2% $Al^{3+}$ doped sample in a zero magnetic field [18].

A common, important feature in many magnetoelectric materials, also seen in $CuFeO_2$ [17] is a magnetic field-induced spin reorientation (spin-flop) transition. Spin reorientation transitions often are accompanied by structural distortions [25] (anomalies in magnetostriction), which are analogous to temperature-induced structural transitions through which electric polarization appears. Spin reorientation transitions are known to show magnetic field-induced structural anomalies in many rare earth orthoferrites, such as $TbFeO_3$ [26], $HoFeO_3$, $DyFeO_3$ [27], $ErFeO_3$ [28] and $TmFeO_3$ [29] which also may be good candidates for magnetoelectric coupling. This stress can lead to lattice-mediated magnetoelectric coupling, because a magnetic field-induced change in the crystalline lattice parameters can stimulate local electric charge displacements, thus inducing magnetic field dependent electric polarization. Indeed, we must emphasize that the $CuFeO_2$ system might provide a new type of magnetic multiferroic because the underlying mechanism has to be different than the inverse Dzialoshinski–Moriya interaction [19–21]. In order to search for the possible existence of magnetodielectric effects in polycrystalline materials, two ceramics of the $CuFeO_2$ and $CuFe_{0.95}Rh_{0.05}O_2$ compounds have been studied. The $Rh^{3+}$ substitution for high spin $Fe^{3+}$was motivated by the closeness of their ionic radii in sixfold coordination ($r_{Rh^{3+}}$= 0.0665 nm, $r_{Fe^{3+}}$=0.0645 nm) which contrasts with the $Al^{3+}$ smaller value ($r_{Al^{3+}}$= 0.0535 nm). In the present paper we report on the

[ǂ] Electronic mail: kundys@.gmail.com (Bohdan Kundys).

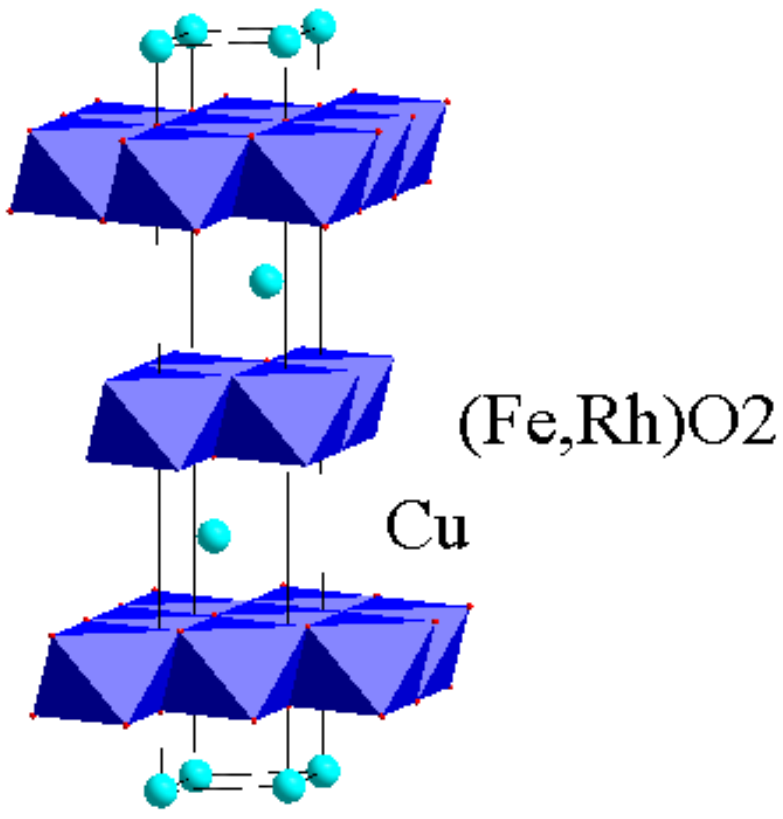

(a)

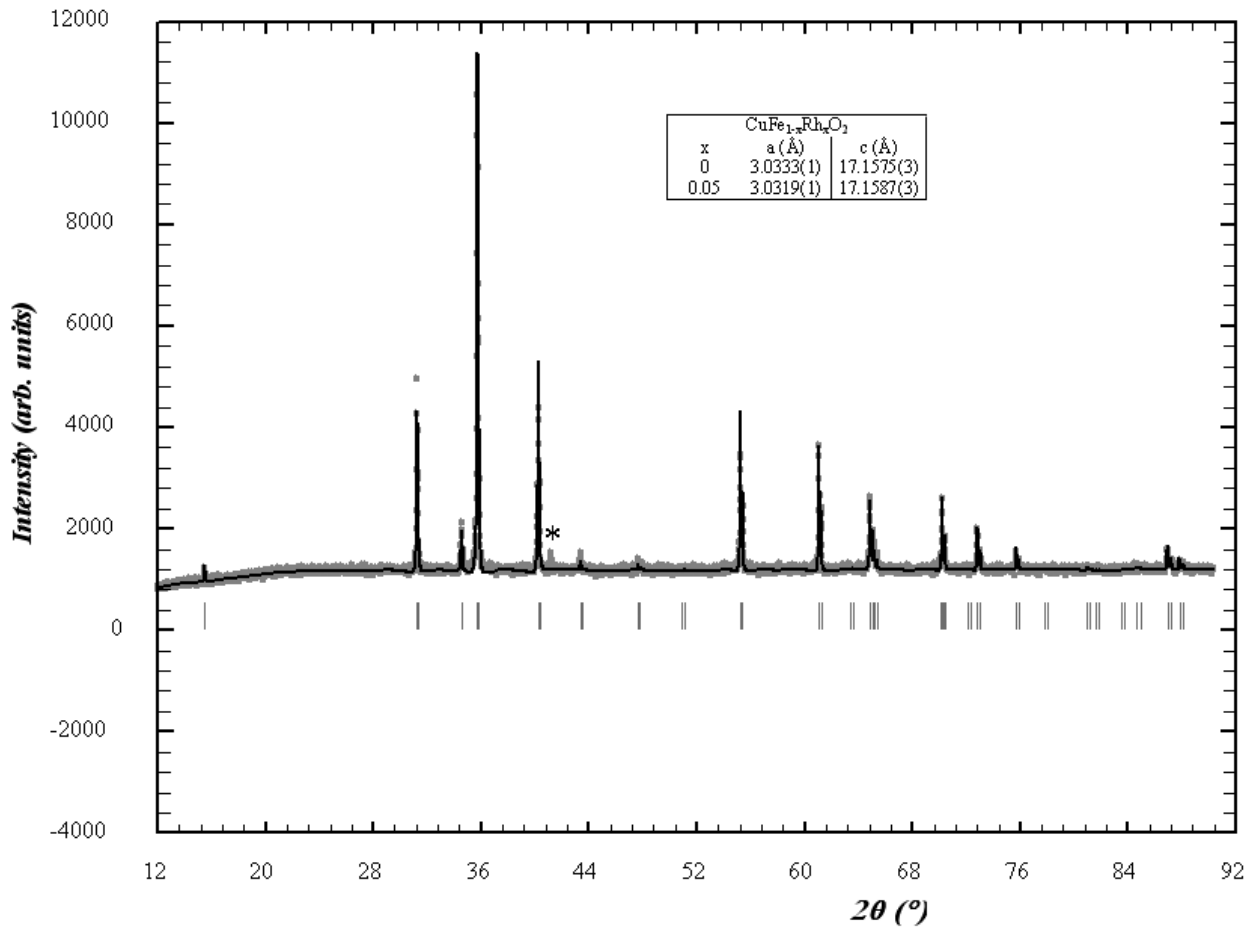

(b)

Fig. 1. (a) Crystalline structure of CuFe0.95Rh0.05O2; (b) experimental powder X-ray diffraction data of CuFe0.95Rh0.05O2 and refined cell parameters (inset).

existence of electric polarization for both $CuFeO_2$ and $CuFe_{0.95}Rh_{0.05}O_2$ even in the absence of external magnetic field. For the pristine compound, these results shed light on the interest of ceramic samples for evidencing magnetoelectric properties. As for applications usage of ceramics has advantages over single crystals, for example in ferroelectric capacitors production, these results demonstrate that multiferroic polycrystals are worth to be studied. Additionally, the search for magnetoelectric multiferroics may be speeded up focusing on ceramics instead of single crystals, which are often very time consuming in preparation.

## 2. Experimental details

The polycrystalline samples were prepared by starting from stoichiometric mixtures of $Cu_2O$, $Fe_2O_3$ and $Rh_2O_3$ which were pressed in bar-shape. Then, about 1g. of bar was set in silica tube. After sealing under primary vacuum, the latter were fired at 1050°C for 12 h. The X-ray powder diffraction patterns (Fig. 1b) of the reacted materials show peaks characteristic of the delafossite structure (Fig. 1a) with the a and c parameters in a R3m hexagonal setting. It must be emphasized that the extra peak marked by an asterisk in the pattern of $CuFeO_2$ (Fig. 1b) corresponds to the presence of FeO traces. Such a result is consistent with the presence of magnetic impurities in a previous study of $CuFeO_2$ when starting from stoichiometric cation amounts [23]. However, as shown later, such an impurity could hardly explain the existence of electric polarization induced by magnetic ordering at the Neel temperature of $CuFeO_2$. The unit cell parameters have been refined from X-ray diffraction data by Rietveld method using the Fullprof software included in the winplot package [30]. The obtained values are in good agreement with those reported in Ref. [23] leading to reliability factors close to 0.07. It is remarkable that the cell parameters for $CuFe_{0.95}Rh_{0.05}O_2$ (Fig. 1b inset) are also very close to that of $CuFeO_2$ as expected from small ionic radius difference between $Rh^{3+}$ and $Fe^{3+}$. In order to check further the purity of the samples, transmission electron microscopy study was made by using a JEOL 2010 microscope equipped with INCA analysis for energy dispersive spectroscopy. The electron diffraction patterns of both samples recorded at room temperature show extinction conditions compatible with the R3m space group in a large majority of samples with FeO oxide as an impurity (less than 5%) like previously detected by X-ray diffraction. For the Rh-doped sample, the concentration determined from energy dispersive analysis coupled to diffraction study for $CuFe_{0.95}Rh_{0.05}O_2$ confirms the presence of rhodium in agreement with starting composition. Taking into account the accuracy of the technique the rhodium content is in the range of $0.04 \leq Rh \leq 0.06$. Additionally, the Curie–Weiss fitting to the linear part of T-dependent reciprocal magnetic susceptibility (PPMS (VSM) magnetometer, zero field cooling 0.3 T) of $CuFe_{0.95}Rh_{0.05}O_2$, yields $\mu_{eff} = 5.87\mu_B$ and $Q_{cw}$=91K, which values are in very good agreement with those reported for $CuFeO_2$ [23]. Magnetodielectric and magnetopolarization measurements were carried out in a PPMS Quantum Design cryostat. The capacitance and dielectric loss were measured using Agilent 4248A RLC meter at different frequencies. Silver pastewas used to make electrical contacts to the sample. The polarization was measured with a Keithley 6517A electrometer. For this the samples were cooled from 20K down to 5K in electric field of about 450 kV/m and in a zero magnetic field. Next the electric field was removed and a time dependence of an electric polarization has been recorded for approximately 2 ks to ensure polarization stability. The temperature dependence of the electric polarization has been recorded thereafter with 1 K/min warming temperature rate. The ferroelectric loop was measured by the homemade dc ferroelectric measurement unit that was used to record ferroelectric loops for $MnWO_4$[31] and $Bi_{0.75}Sr_{0.25}FeO_{3-d}$[32] magnetic multiferroics.

## 3. Results and discussions

The temperature dependent measurements are given in Fig. 2 at zero applied magnetic field for both doped and undoped sample. The magnetization curve and dielectric permittivity of polycrystalline $CuFeO_2$ agree well with previous report for monocrystalline sample (Fig. 2a). The magnetization curve exhibits a well-defined anomaly with a maximum near 12 K that is also correlated with a dielectric peak (Fig. 2b). The broad dielectric anomaly starting at about 15 K with a maximum near 12 K can therefore be associated to the formations of antiferromagnetic phases, from paramagnetic to incommensurate collinear ($T_{N1}$=15K) and then to a commensurate collinear four lattice antiferromagnetic structure ($T_{N2}$=12K) [14]. However Rh-doping affects both dielectric permittivity and magnetization curves (Fig. 2a and b). In particular, the Neel temperature is shifted towards lower temperature region and the magnetization peak corresponding to TN2 is suppressed (Fig. 2a). While no peak is seen in the temperature dependence of magnetization near 9 K (Fig. 2a), the anomaly can still be evidenced in its derivative plot (see in the inset of Fig. 2a). At the same temperature a clear peak in the dielectric permittivity is induced by $Rh^{3+}$ substitution for $Fe^{3+}$ (Fig. 2b). Furthermore, a doping induced electric polarization effect is seen for $CuFe_{0.95}Rh_{0.05}O_2$ (Fig. 2c) characterized by an abrupt increase of the electric polarization below that temperature (≈9 K) reaching the maximum near 8.3 K. This is in accordance with a previous paper reporting on Al doping induced electric polarization in monocrystalline sample of $CuFeO_2$ [18]. In order to verify the nature of polarization effect, the polarization of the undoped $CuFeO_2$ sample has been also measured using the same electric field cooling procedure as was applied for $CuFe_{0.95}Rh_{0.05}O_2$ (Fig. 2c). It also reveals a small electric polarization for pure $CuFeO_2$, whose maximum at w12 K is correlated with anomalies in both magnetization (Fig. 2a) and dielectric permittivity (Fig. 2b). This provides new data for the physics of $CuFeO_2$, since the electric polarization was reported in single crystal in the same temperature region but only under magnetic field application [17]. The existence of electric polarization in the polycrystalline pristine sample in zero external magnetic field indicates that spontaneous electric polarization may exist along other crystallographic direction than those already tested in single crystals. Therefore this suggests that an electric polarization flop between different directions may be induced by magnetic field application in the single crystal of $CuFeO_2$. It was also checked by measurements at different frequencies that there is practically no frequency dependence in the position of dielectric anomalies (Fig. 2b).

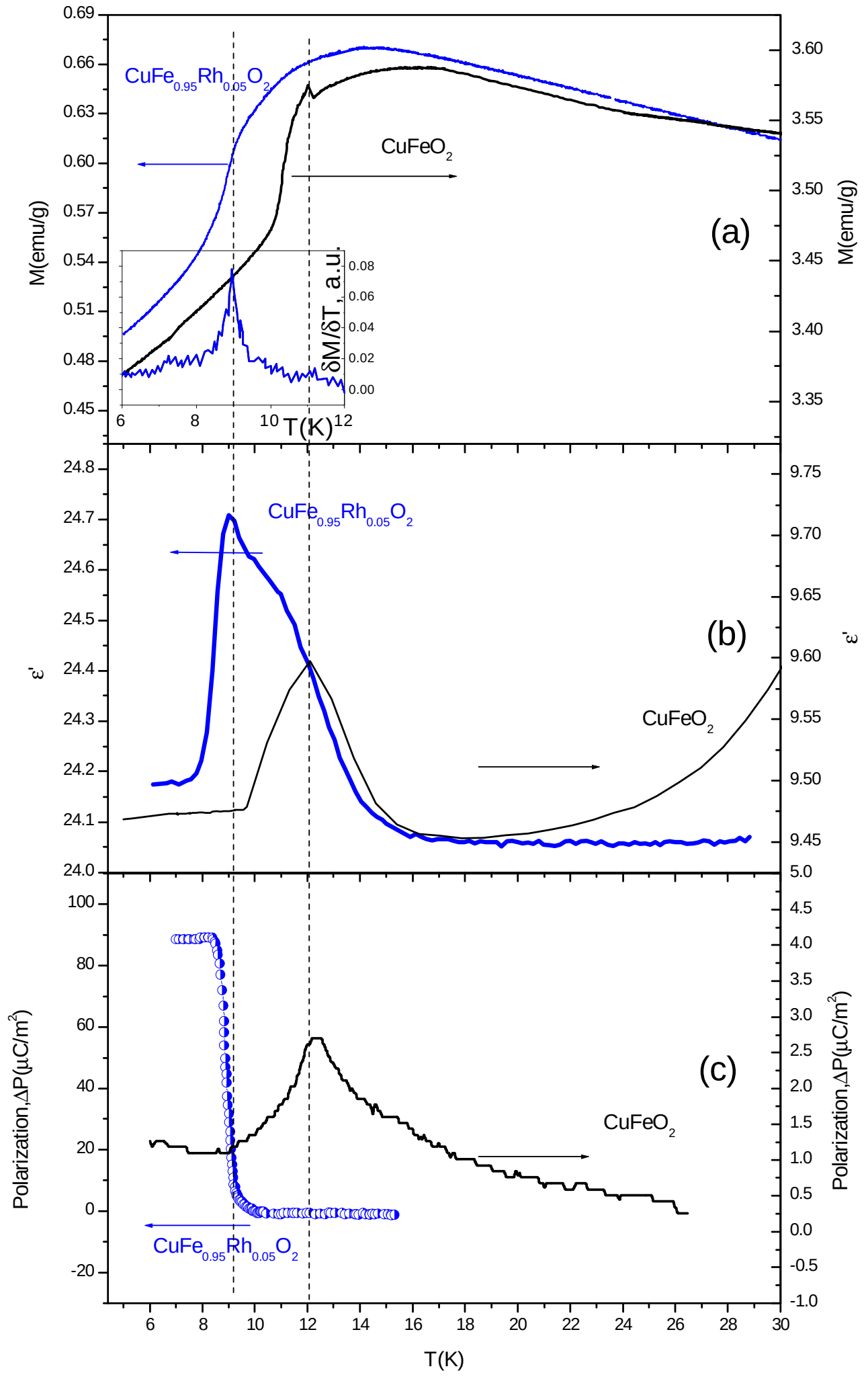

Fig. 2. ZFC (0.3 T) magnetization (a), dielectric permittivity measured on cooling (b) and electric polarization (c) of CuFeO2 and $CuFe_{0.95}Rh_{0.05}O_2$ ceramics as a function of temperature. The dM/dT derivative curve for the Rh-doped compound is given in the inset of Fig. 2(a)

The existence of electric polarization in both samples related to magnetic transition at $T_{N2}$ implies the existence of intrinsic magnetoelectric coupling in the present polycrystalline samples. Consequently, for $CuFe_{0.95}Rh_{0.05}O_2$ magnetodielectric effect has been examined at several temperatures (Fig. 3a) below the characteristic temperature (T≈9K) of the dielectric anomaly. Dielectric permittivity measurements reveal symmetrical branches for positive and negative applied magnetic field with a hysteresis presenting a butterfly shape. They also show a magnetic field-induced maximum ($\Delta\varepsilon'/\varepsilon(0)$ =0.015 at 8.8 T) of the dielectric permittivity for T <8.3 K, which is known to be a good indication of polar–non polar transitions [8,9,31]. Coming from the low temperature (6 K) as temperature increases the peak shifts to lower magnetic fields and beyond 8.3 K the magnetodielectric curves start to change their shape. The change in sign is observed for 9K with a minimum of magnetodielectric coupling of

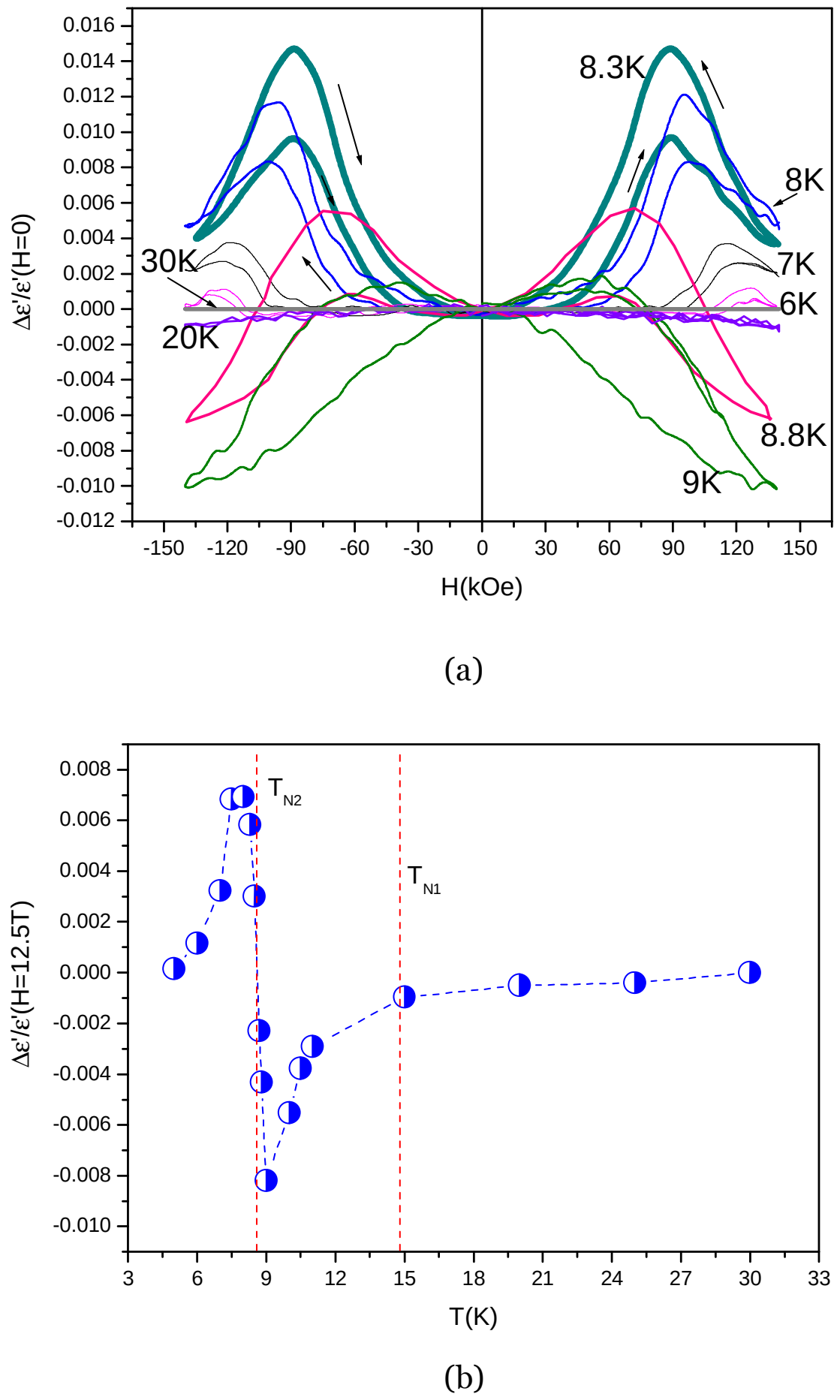

Fig. 3. Relative dielectric permittivity at 100 kHz as a function of magnetic field taken at different temperatures for $CuFe_{0.95}Rh_{0.05}O_2$. The arrows indicate increasing and decreasing magnetic fields (a). Temperature dependence of the relative dielectric permittivity at 100 kHz taken at 12.5 T for $CuFe_{0.95}Rh_{0.05}O_2$ (b).

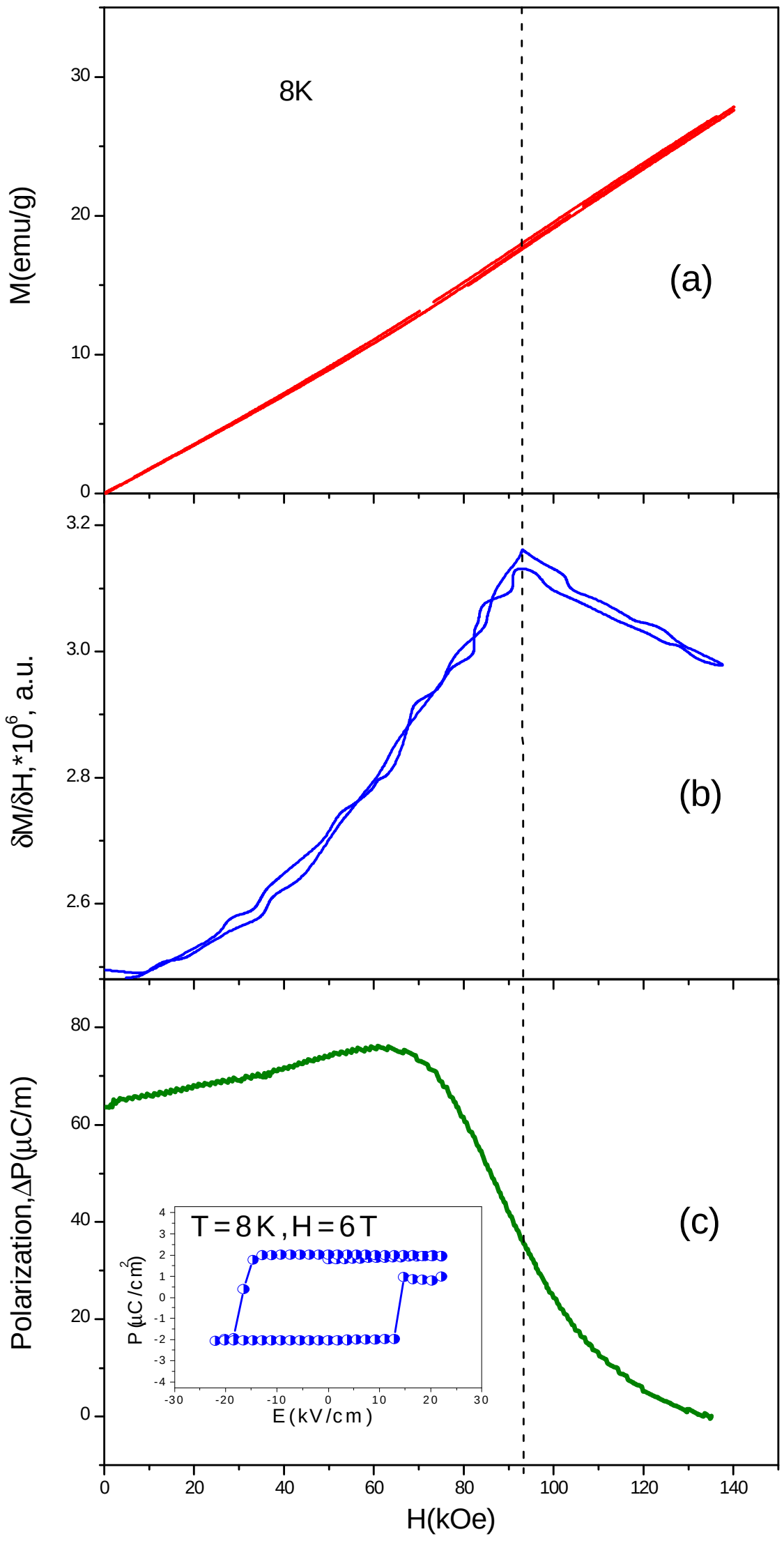

Fig. 4. Data for $CuFe_{0.95}Rh_{0.05}O_2$. Magnetic field dependence of magnetization (a), derivative of magnetization (b) and electric polarization (c) taken at T= 8 K after cooling from 20 K in 340 kV/m. Inset shows ferroelectric loop taken at 6 T.

$\Delta\varepsilon'/\varepsilon(0) =$ -0.010 at magnetic field of 14 T. While the shape of the dielectric permittivity versus magnetic field remains similar for electrically polar region, the magnitude of the magnetodielectric effect and the magnetic field position of the observed maximum are strongly affected by temperature. As is seen in Fig. 3a as temperature increases, the magnitude of the peak and its characteristic magnetic field increases and decreases towards zero respectively. In the paraelectric antiferromagnetic temperature region, the magnitude of magnetodielectric effect is now different and no obvious peak is seen. With further warming, zero magnetodielectric effect is observed for T=30 K. The relative magnitude of the magnetodielectric effect taken at given magnetic field (12.5 T) reveals a discontinuity (Fig. 3b). This behavior strongly suggests that the mechanism of the magnetodielectric coupling for noncollinear antiferromagnetic region (T<8.3K Fig. 2a and c) differs significantly from that of paraelectric collinear antiferromagnetic region. A similar behavior, with a magnetoelectric effect which increases as temperature approaches the transition as shown in Fig. 3b (see electrically polar region T$\leq$8.3 K), was already observed for $Cr_2O_3$ [33] and recently for $YBaCuFeO_5$ [34]. This effect was analyzed by a simple phenomenological model [35]. However, the observed discontinuity at the transition brings new information in the magnetodielectric properties of solids and evidences different nature of magnetodielectric coupling in the two different antiferromagnetic states. Moreover, this experimental fact confirms that magnetodielectric effects may be present even in nonpolar antiferromagnetic region. The relative magnitude of the magnetodielectric effect is found to be frequency independent (not shown) and the dielectric losses are found to be of the order of $10^{-3}$ or less. This behavior strongly implies an intrinsic origin of the magnetodielectric coupling between magnetic and electric orders in the sample. This is confirmed by the electric polarization (P) measured at T= 8 K showing a clear magnetic field dependence (Fig. 4c). As the magnetic field rise. This can be associated to the magnetic field dependence of the magnetization (Fig. 4a) which shows a change in its slope (see the differential plot in Fig. 4b) at the same magnetic field region (≈9.3T) where the peak in dielectric permittivity is observed (Fig. 3a). The fact that electric polarization is suppressed by an external magnetic field application implies that magnetic field changes electrically favorable spin configuration and induces polar–paraelectric transition. Additionally the electric polarization at 6 T (maximum on the P(H) curve Fig. 4c) may be also switched by rather high electric field (≈16 kV/cm) and thus demonstrates ferroelectricity (with $P_r \approx 2 \mu C/cm^2$) at 8 K (Fig. 4c inset). However, ferroelectric fatigue effect is probably responsible for the observed imperfection of ferroelectric loop. The experimental fact that no ferroelectric loop was observed beyond 9.3T also supports magnetic field-induced ferroelectric–paraelectric transition. Based on both dielectric and magnetic data a (T, H) phase diagram of the Rh-doped sample is obtained and given in Fig. 5. For the latter, as the nature of the magnetic phases is unknown, the properties similarity for the present $CuFe_{0.95}Rh_{0.05}O_2$ polycrystalline sample and the Al-doped crystal [17,20], leads us to propose that Rh-doping extends the region of the polar incommensurate noncollinear phase as compared to $CuFeO_2$.

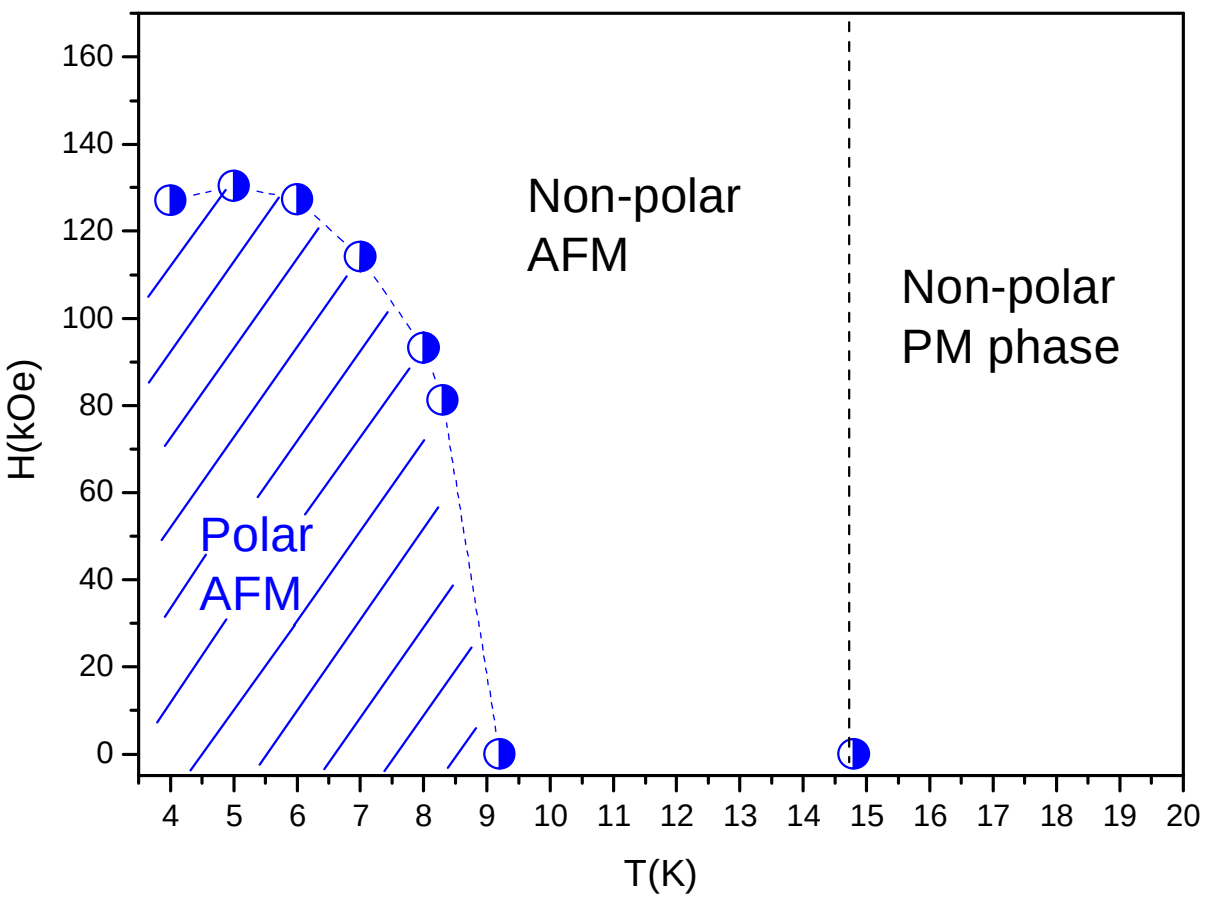

Fig. 5. Magnetoelectric phase diagram of $CuFe_{0.95}Rh_{0.05}O_2$.

## 4. Conclusions

The anomaly in dielectric permittivity associated with the noncollinear antiferromagnetic phase that has been reported previously for $CuFeO_2$ and Al-doped $CuFeO_2$ monocrystals, has been also found in polycrystalls $CuFe_{0.95}Rh_{0.05}O_2$. Without an externally applied magnetic field, the evidence for electric polarization in the polycrystalline $CuFeO_2$ contrasts with the lack of observed polarization in the single crystal underscoring the potential advantage of using ceramic samples to search for new multiferroics. This phenomenon strongly supports the fact that even in materials that have very anisotropic structures such as delafossites, the averaging of the electric polarization that is generated by measurements that are performed on ceramics, i.e. over the microcrystals assembly with an isotropic distribution is not a redhibitory limitation, but can even provide an advantage over the crystals. The study of Rh-doped ceramics reveals significant changes in the magnetoelectric phase diagramas the Neel temperature shifts toward lower temperatures. The substitution also induces an incremental shape for the P(T) curve, wherein the electric polarization increases abruptly near 9 K. This effect can be compared with the induced change of the background magnetic state that has been reported for Al-doped crystals. Finally, the similarity of the effects induced by either $Al^{3+}$or $Rh^{3+}$ demonstrates that the ionic size of the substituted cations is not so important but rather that S =0 electronic configuration is highly favorable for doping induced electric polarization effect. More importantly, the fact that polarization is observed for 5% $Rh^{3+}$ concentration in zero magnetic field indicates the possible advantage of the $Rh^{3+}$ doping over the $Al^{3+}$ one. The latter is reported to induce incommensurate magnetic structure only in the doping range of $0.014 \leq Al^{3+} \leq 0.03$ [15]. In that respect a complete study of the $CuFe_{1-x}Rh_xO_2$ series by comparing crystals and ceramics could be of basic interest. In the future, the present study should motivate the screening of frustrated magnetic materials in the search for new magnetic multiferroics.